\documentclass[twocolumn,showpacs,superscriptaddress,pra]{revtex4-1}

\usepackage{amsmath, amsthm, amssymb}
\usepackage{graphicx}
\usepackage{dcolumn}
\usepackage{bm}
\usepackage{color}

\newcommand{\ket}[1]{\left | #1 \right \rangle}
\newcommand{\bra}[1]{\left \langle #1   \right |}

\newcommand{\Tr}{\operatorname{Tr}}

\newcommand{\comment}[1]{}

\newcommand{\ave}[1]{\langle#1\rangle}

\newcommand{\ba}{\begin{eqnarray}}
\newcommand{\be}{\begin{equation}}
\newcommand{\ee}{\end{equation}}

\newcommand{\ea}{\end{eqnarray}}
\newcommand{\ban}{\begin{eqnarray*}}
\newcommand{\ean}{\end{eqnarray*}}

\newcommand{\ie}{{\it{i.e.}}}

\begin{document}
\title{Detecting nonlocality of noisy multipartite states with the CHSH inequality}
\author{Rafael Chaves}
\affiliation{Institute for Physics, University of Freiburg, Rheinstrasse 10, D-79104 Freiburg, Germany}
\author{Antonio Ac\'in}
\affiliation{ICFO-Institut de Ci\`encies Fot\`oniques, Mediterranean
Technology Park, 08860 Castelldefels (Barcelona), Spain}
\author{Leandro Aolita}
\affiliation{Dahlem Center for Complex Quantum Systems, Freie Universit\"{a}t Berlin, Berlin, Germany}
\author{Daniel Cavalcanti}
\affiliation{ICFO-Institut de Ci\`encies Fot\`oniques, Mediterranean
Technology Park, 08860 Castelldefels (Barcelona), Spain}

\begin{abstract}
The Clauser-Horne-Shimony-Holt inequality was originally proposed as a Bell inequality to detect nonlocality in bipartite systems. However, it can also be used to certify the nonlocality of multipartite quantum states. We apply this to study the nonlocality of multipartite Greenberger-Horne-Zeilinger, W and graph states under local decoherence processes. We derive lower bounds on the critical local-noise strength tolerated by the states before becoming local. In addition, for the whole noisy dynamics, we derive lower bounds on the corresponding nonlocal content for the three classes of states. All the bounds presented can be calculated efficiently and, in some cases, provide significantly tighter estimates than with any other known method. For example, they reveal that $N$-qubit GHZ states undergoing local dephasing are, for all $N$, nonlocal throughout all the dephasing dynamics.
\end{abstract}

\maketitle
\section{Introduction}
Non-locality refers to correlations between the measurement results of distant systems that cannot be explained by
local hidden-variable (LHV) models~\cite{Bell1964,NLReview2013}. The correlations consistent with a LHV model necessarily satisfy a set of linear constraints known as {\it Bell inequalities}~\cite{NLReview2013}, which can be experimentally tested. Thus, the violation of any Bell inequality reveals the presence of non-locality. In addition, apart from a fundamental issue, the detection of nonlocal correlations is also of practical relevance. First, the violation of a Bell inequality is a device-independent entanglement witness, i.e. it allows one to certify entanglement in situations where the sources and measurements implemented are totally unknown \cite{Bancal2011,NLReview2013}.  Second, the efficacy at solving  information-theoretic tasks such as communication complexity problems~\cite{Buhrman2010}, device-independent quantum key distribution~\cite{Acin2007,Barrett2005,Acin2006} and randomness extraction~\cite{Pironio2010,Colbeck2011} or amplification \cite{Colbeck12,Gallego13,Ramanathan13} relies on the presence of nonlocality. Experimentally-friendly ways to extract nonlocal correlations from quantum states appears thus highly desirable.

The simplest way to do this, in the case of two parts with two dichotomic measurements each, is through the CHSH inequality \cite{Clauser1969}
\be
CHSH\equiv\ave{a_0 b_0}+\ave{a_0b_1}+\ave{a_1b_0}-\ave{a_1b_1}\leq 2,
\label{CHSH}
\ee
where $a_x=\pm1$ and $b_y=\pm1$ are the outcomes of measurement settings labeled by $x=\{0,1\}$ and $y=\{0,1\}$ for Alice and Bob, respectively, and $\ave{a_x b_y}=p(a=b|xy)-p(a\neq b|xy)$ stands for the statistical average of $a_x b_y$. In quantum mechanics these averages can be expressed by $\ave{a_x b_y}\equiv\Tr[\hat{A}_x\otimes\hat{B}_y \rho]$, where $\hat{A}_x$ and $\hat{B}_y$ are Hermitian observables with eigenvalues $\pm1$ and $\rho$ a quantum state. The CHSH inequality (and its symmetries) is the only relevant Bell inequality in the bipartite scenario with two dichotomic measurements \cite{Fine1982}, i.e. it can tightly capture all non-local correlations. Furthermore, for two-qubit quantum states, CHSH violation can be immediately checked via the necessary and sufficient condition found in~\cite{Horodecki1995}.

In the multipartite scenario, however, the situation changes drastically. For instance, already for the modest case of three parts applying two dichotomic measurements each, there are 46 inequivalent classes of non-trivial and tight Bell inequalities \cite{Sliwa2003}. In general, the efficiency in the characterization of nonlocality as the number of parts, measurements or outcomes increases becomes a major issue. In fact, deciding the compatibility of a given probability distribution with LHV models is known to be an NP-complete problem \cite{Pitowsky1989,Pitowsky1991}.

In this paper, we study the nonlocality of genuinely multipartite $N$-qubit Greenberger-Horne-Zeilinger (GHZ), W and graph states under local decoherence processes described by Pauli channels. We derive lower bounds on the critical local-noise strength tolerated by the states before becoming local, in a similar spirit as in \cite{Laskowski2010}. In addition, for each noise strength, we derive lower bounds on the nonlocal content \cite{Elitzur1992} of the correlations on the three classes of states. The bounds we derive are based on the CHSH violation of two out of the $N$ qubits conditioned on a measurement outcome of all other $N-2$ qubits \cite{Popescu1992}, and can therefore be calculated efficiently. As a matter of fact, we show that in some cases, such as with GHZ states under transversal local dephasing (bit-flip noise), the bounds obtained are even $N$-independent. Furthermore, we show that the estimates given by these bounds are (in some cases exponentially) tighter than those given by any other known method.

In Sec. \ref{sec:SNF} we introduced the different classes of states, the noise models and figures of merit for nonlocality to be used in this paper. In Sec. \ref{sec:method} we describe the general method that is applied in Sec. \ref{Results} to derive, respectively, lower bounds on the critical noise strength and the nonlocal content. In \ref{sec:conc} we present a summary of the results while some technical results about graph states Bell inequalities are relegated to the Appendix.

\section{States, noise models and figures of merit}
\label{sec:SNF}
In this section, we introduce basic notation, define the states studied, the noise channels considered and the figures of merit we use to assess the non-locality of noisy states.
\subsection{States under scrutiny}

We consider three paradigmatic families of genuinely multipartite $N$-qubit quantum states:
\begin{itemize}
\item  GHZ states \cite{GHZ}
\begin{equation}
\label{GHZ}
\ket{GHZ_N} \doteq \frac{1}{\sqrt{2}} ( \ket{0}^{\otimes N} + \ket{1}^{\otimes N});
\end{equation}

\item   W states \cite{Dur2000}
\begin{equation}
\left\vert W_N\right\rangle \doteq\frac{1}{\sqrt{N}}\left(  \left\vert
0\ldots01\right\rangle +\left\vert
0\ldots10\right\rangle +\ldots+\left\vert 10\ldots0\right\rangle \right);
\end{equation}

\item Graph states \cite{Hein2004,Hein2006}. A graph-state $\ket{G_{\bf{0}}}$ is associated to an $N$-vertex mathematical graph $G$, whose geometry is determined by a set $\mathcal{E}$ of edges $\{i,j\}$ indicating which vertices $i$ and $j$ are connected, for  $1\leq i,j\leq N$. More precisely,
\begin{equation}
\label{graphdef}
\ket{G_{\bf{0}}}\doteq CZ_{\mathcal{E}}\ket{+}^{\otimes N},
\end{equation}
being $\ket{+}\doteq(\ket{0}+\ket{1})/\sqrt{2}$ and $CZ_{\mathcal{E}}\doteq\prod_{\{i,j\} \in \mathcal{E}}CZ_{i,j}$, where $CZ_{i,j}\doteq e^{(Z_i-\openone_i)\otimes(Z_j-\openone_j)/4}\otimes\openone_{\overline{i,j}}$ is the maximally entangling controlled-Z gate non-trivially acting on qubits $i$ and $j$, with $Z_i$ and $Z_j$ the third Pauli operators on qubits $i$ and $j$, respectively, and $\openone_i$, $\openone_j$, and $\openone_{\overline{i,j}}$ the identity operators on qubits $i$, $j$, and all but $i$ and $j$, respectively, for any $1\leq i,j\leq N$.

\end{itemize}

\subsection{Decoherence models}
As noise models we consider local Pauli channels of the form
\begin{equation}
\Lambda(\rho_0) \doteq \sum_{i=0}^3 p_i \sigma_i \rho_0\sigma_i.
\label{Paulimap}
\end{equation}
Here, $\rho_0$ is any initial state and $\Lambda$ is a single-qubit Pauli channel.  $\sigma_0\doteq\openone$,  and $\sigma_1\doteq X$, $\sigma_2\doteq Y$ and $\sigma_3\doteq Z$ refer to the usual Pauli operators. The coefficients  $p_i$ satisfy the relationship $p_0=(1-p/2)$, $p_1=\alpha_1 p/2$, $p_2=\alpha_2 p/2$, $p_3=\alpha_3 p/2$, with $\alpha_1 + \alpha_2 + \alpha_3 \!=\!1$; so that the total noise strength $0\leq p \leq 1$ is distributed along the three Bloch-sphere directions according to $\alpha_1$, $\alpha_2$ and $\alpha_3$. For example, the case $\alpha_1=\alpha_2=0$  describes dephasing along the direction $z$ of the Bloch sphere (also known as phase-flip channel). Analogously, $\alpha_2=\alpha_3=0$ describes dephasing along the transversal direction $x$ (bit-flip channel). We consider joint evolutions given by independent and identical channels on all qubits:
\begin{equation}
\rho_{p}=\Lambda^{\otimes N}(\rho_{0}).
\label{JointPaulimap}
\end{equation}

\subsection{Figures of merit}
To assess the non-local correlations in quantum states, we focus mainly  in two quantities. The first one is the critical noise strength $p_c$ beyond which on no non-locality can be extracted  \cite{Laskowski2010}.
We refer to $p_c$ as the noise threshold and in the following we compute a lower bound to it.

The second one is the amount of nonlocality for each noise strength $p$, which we quantify through the EPR2 decomposition \cite{Elitzur1992}. Any joint-probability distribution $P$, characterising the correlations of some Bell experiment, can be decomposed into convex mixture of a local part $P_{L}$ and a general non-local (no-signalling)  part $P_{NL}$ as $P = (1-p_{NL}) P_{L} + p_{NL}P_{NL}$, with $0\leq p_{NL}\leq1$.
The minimal non-local  weight over all such decompositions,
\begin{eqnarray}
\label{ptilde}
\tilde{p}_{NL}&\doteq&\min_{P_{L},P_{NL}}p_{NL}.
\end{eqnarray}
defines the \emph{nonlocal content} of $P$, which provides a natural quantifyer of the non-locality in $P$. In turn, we define the non-local content of a quantum state as the maximum non-local content of correlations over all possible Bell experiments with the state.

It turns out that the violation of any Bell inequality yields a non-trivial lower bound to $\tilde{p}_{NL}$ \cite{Barrett2006}. For any (linear) Bell inequality $\mathcal{I}\leq\mathcal{I}^L$, with $\mathcal{I}^L$  the local bound, it is
\begin{equation}
\label{lowerbound}
    \tilde{p}_{NL}\geq
\frac{\mathcal{I}(P)-\mathcal{I}^{L}}{\mathcal{I}^{NL}-\mathcal{I}^{L}},
\end{equation}
where $\mathcal{I}^{NL}$ is the maximum Bell value $\mathcal{I}$ over all arbitrary non-signalling correlations.
\section{The method}
\label{sec:method}

We will consider a scenario where $N$ parties share a multipartite state and perform local measurements on it. Two of the parties apply two dichotomic measurements, labeled again by $x=\{0,1\}$ and $y=\{0,1\}$, with possible outcomes $a_x=\pm1$ and $b_y=\pm1$, respectively.  The other $N-2$ parties apply only one dichotomic measurement each. We will denote the outcomes of these $N-2$ measurements by a $N-2$ component vector $\textbf{c}=(\pm1,\ldots,\pm1)$. The Bell inequality that we will consider is given by (see Appendix B of ref. \cite{Cavalcanti2011})
\be
CHSH_{\bf c}\equiv\ave{a_0 b_0}_{{\bf c}}+\ave{a_0b_1}_{{\bf c}}+\ave{a_1b_0}_{{\bf c}}-\ave{a_1b_1}_{{\bf c}}- 2p({{\bf c}})\leq 0,
\label{conditioned CHSH}
\ee
where $\ave{a_x b_y}_{{\bf c}}=[p(a=b|x,y,{{\bf c}})-p(a\neq b|x,y,{{\bf c}})]p({{\bf c}})$. Notice that this inequality is simply the CHSH inequality calculated with the conditional probability distribution for the two parties given that the other $N-2$ parties get the particular outcome ${\bf c}$.

Proof of the validity of Bell inequality \eqref{conditioned CHSH}: We need to show that all the local deterministic probability distributions, \ie~ those assigning definite outcomes for each measurement, satisfy it. For the local deterministic distributions for which $p(\textbf{c})=1$, inequality \eqref{conditioned CHSH} becomes the standard CHSH inequality \eqref{CHSH}, while for the local deterministic strategies such that $p(\textbf{c})=0$  it simply reads $0\leq0$. $\square$

Thus, in order to detect nonlocality in a given $N$-partite state $\rho$ through the inequality \eqref{conditioned CHSH} we need to find appropriate local measurements on $N-2$ parts that project the remaining two parts into a bipartite state violating the CHSH inequality \cite{Popescu1992}.
At this point it is worth emphasising that the conditioning (or post-selection) used in the present Bell test does not open any loophole. The reason is that it is done only on the outcomes of  the $N-2$ parts which are space-like separated from the two parties involved in the CHSH test. In this way, it could simply be seen as a heralded preparation of a nonlocal state by $N-2$ parties. Moreover, this method has already proven very useful in other contexts. For instance, it has been applied to prove that every multipartite pure entangled state is nonlocal  \cite{Popescu1992}, and to demonstrate super-activation of nonlocality in quantum networks \cite{Cavalcanti2011,Cavalcanti2012,Klobus2012}.

Here, given projective measurements on the first $N-2$ qubits, we will test the nonlocality of the resulting two-qubit state through the necessary and sufficient condition for CHSH violation discovered in Ref. \cite{Horodecki1995}: The maximum CHSH value achievable by an arbitrary two-qubit state $\rho$ is
\begin{equation}\label{chsh criterion}
CHSH=2\mathcal{M}_{\text{CHSH}}(\rho)=2\sqrt{t_{11}^{2}+t_{22}^{2}},
\end{equation}
being $t_{11}^{2}$ and $t_{22}^{2}$ the two largest eigenvalues of $\mathcal{T}^{\dag}\mathcal{T}$, with $\mathcal{T}_{i,j}=\rm{tr}\left[  \left(  \sigma_{i}\otimes\sigma_{j}\right)\rho\right]$, where $\sigma_{i}\otimes\sigma_{j}$ refers to the product of the $i$-th and $j$-th Pauli operators on the two remaining qubits, for $1\leq i,j\leq 3$. So, $\rho$ violates the CHSH inequality if, and only if,  $\mathcal{M}_{\text{CHSH}}(\rho)>1$.

As long as the probability $p({{\bf c}})$ is greater than zero, its exact value does not affect the critical noise thresholds. Note, however, that using inequality \eqref{conditioned CHSH} the lower bound for the nonlocal content will unavoidably depend on $p({{\bf c}})$, namely,
\begin{equation}
\label{neweq}
\tilde{p}_{NL}\geq \frac{[\text{CHSH}(\rho)-2]p({\bf c})}{2}.
\end{equation}
 For the states we consider in this paper, $p({\bf c})$ will typically decay exponentially with the number of qubits $N$, also leading to a exponentially decaying lower bound. In order to circumvent that and still get non trivial lower bounds for the nonlocal content we proceed as follows.

For all the states we consider (with the exception of the W state considered in Sec. \ref{subsec:wstate}) all possible $2^{N-2}$ measurements outcomes lead to only $2$ possible projected two-qubit states that, furthermore, are equivalent up to local unitaries. Let us call these projections as events $1$ and $2$, the two respective projected states by $\rho_1$ and $\rho_2$, and $p(1)$ and $p(2)=1-p(1)$ the probabilities of events $1$ and $2$.

We can then define Bell inequalities, similar to \eqref{conditioned CHSH}, to events $1$ and $2$ as
\be
\label{CHSH1}
CHSH_1\equiv\ave{a_0 b_0}_{1}+\ave{a_0b_1}_{1}+\ave{a_1b_0}_{1}-\ave{a_1b_1}_{1}-2p(1)\leq0,
\ee
and
\be
\label{CHSH2}
CHSH_2\equiv\ave{a_0 b_0}_{2}+\ave{a_0b_1}_{2}-\ave{a_1b_0}_{2}+\ave{a_1b_1}_{2}-2p(2)\leq0,
\ee
with $\ave{a_x b_y}_{i}=[p(a=b|x,y,i)-p(a\neq b|x,y,i)] p_i$. Finally we use these inequalities to define the following one:
\be\label{sumCHSH}
CHSH_1+CHSH_2\leq0.
\ee
For most of the states we will consider here, we can find measurements ${ A_0,A_1,B_0,B_1}$ that will lead to $p(1)=p(2)$ and $CHSH_1=CHSH_2$. This, in turn, will imply that the lower bound for the nonlocal content will be independent of the projection probabilities and simply given by $\tilde{p}_{NL}\geq \mathcal{M}_{\text{CHSH}}(\rho)-1$.

Finally notice that different Bell inequalities, conditioned on outputs of $N-2$ parties, could be similarly used. However in the case that the two parties testing the Bell inequality have two binary inputs it is sufficient to consider the CHSH inequality.


\section{Nonlocality threshold and nonlocal content of noisy states}
\label{Results}
In this section we show how the multipartite CHSH method can be used to calculate the critical noise strength tolerated by the noisy state before becoming local. We also compute, for the entire noisy dynamics, lower bounds for the nonlocal content of the states. These lower bounds can be significantly better than the ones obtained via known multipartite inequalities.

\subsection{Noisy GHZ state}
\label{sec:pcriticalsec}
We begin considering GHZ states. In particular, for parallel dephasing, we show that GHZ states of any number of qubits are nonlocal throughout all the noisy dynamics, a result that cannot be achieved by any other known multipartite inequality consisting exclusively of full-correlators.

\subsubsection{Parallel dephasing}

We consider first the detection of nonlocality for the GHZ state~\eqref{GHZ} undergoing independent dephasing along the $Z$ direction. The resulting noisy GHZ state $\rho_N^{z}$ can be expressed as \cite{Aolita2008,Aolita2009}
\begin{equation}
\rho_N^{z} = (1-p)^{N} \ket{GHZ_N}\bra{GHZ_N}
+ \left( 1- (1-p)^{N} \right) \tilde{\varrho}_N^{z},
\label{GHZdephased}
\end{equation}
with $\tilde{\rho}_N^{z}=\big(\ket{0}\bra{0}^{\otimes N} + \ket{1}\bra{1}^{\otimes N}\big)/2$.

We compare the multipartite CHSH method with the Werner-Wolf-Weinfurter-Zukowski-Brukner (W${}^3$ZB) inequalities \cite{Werner2001,Weinfurter2001,Zukowski2002}. These encompass all the $2^{2^N}$ tight, linear, full-correlator Bell inequalities in the $N$-partite scenario where each party makes two dichotomic measurements. In particular, of special relevance here is the Mermin-Klyshko (MK) inequality \cite{Mermin1990,Belinskii1993,Scarani2001}, which is a particular case of the W${}^3$ZB family. The MK inequality is the two-setting correlator Bell inequality with the largest violation in quantum theory \cite{Werner2001}, with an exponential maximal violation $2^{(N-1)/2}$ (the local bound of the MK inequality is given by $1$), achieved with the GHZ state for $X$ and $Y$ measurements.

The maximal MK violation for $\rho_N^{z}$ can be straightforwardly calculated \cite{Chaves2012b} for the case of $N$ odd, to which we restrict for simplicity of notation. It is given by $2^{(N-1)/2}(1-p)^N$ and is also attained with $X$ and $Y$ measurements. This yields in turn the noise threshold $p^{z}_{c}=1-1/\sqrt{2}^{\left( N-1\right) /N}$, which is tighter than that given by any other known multipartite inequality consisting exclusively of full-correlators.

We next show that the CHSH method renders $p_c^z=1$ for all $N$. Consider local $X$ measurements on the first $N-2$ qubits of \eqref{GHZdephased} (Numerical optimization up to $N=5$ shows that these measurements are optimal, that is, they maximize the CHSH violation of the remaining two-qubit state. See \cite{noteoptimal} for further details). We consider explicitly the situation where all $N-2$ parties obtain the eigenvalue 1, corresponding to the eigenstate $\ket{+}$. However, for any other outcome the treatment would be equivalent, except for a local-unitary relabelling of the projected states. This local-unitary equivalence will be explicitly used later on in order to derive lower bounds to the nonlocal content. The projected two-qubit state conditioned on the $N-2$ measurement outcomes obtained is
\begin{equation}
\rho_2^{z} = (1-p)^N\ket{GHZ_{2}}\bra{GHZ_{2}}+(1-(1-p)^N) \tilde{\rho}_2^{z}.
\label{GHZdephasedprojected}
\end{equation}
Computing \eqref{chsh criterion} for this state gives $\mathcal{M}^{z}_{\text{CHSH}}=\sqrt{1+(1-p)^{2N}}$, which is greater than one for all $p<1$. Thus, GHZ states of arbitrary $N$ subject to independent parallel dephasing are non-local for any amount of dephasing $p<1$. We stress that such high noise threshold cannot be detected by any other known multipartite inequality consisting exclusively of full-correlators.

Interestingly, for $N=3$ this result can be made even stronger, since the CHSH method is able to detect the nonlocality in a region where \emph{any} full-correlator inequality would fail. For the state \eqref{GHZdephased} and $N$ odd it is not difficult to see that only the components of the projective measurement lying in the equatorial plane give a non-null contribution for full-correlators. For example for $N=3$ and the observable $O=\frac{(X+Z)}{\sqrt{2}} \otimes X \otimes X$ we have that $\rm{tr}(O \rho_N^{z})=\rm{tr}((X\otimes X \otimes X)((1/\sqrt{2})\rho_N^{z}+(1-1/\sqrt{2})\openone))$, and so in this sense it is sufficient to only consider equatorial measurements. On the other hand, it follows from Ref. \cite{BG11} that for $p\geq1/2$ any equatorial measurements on the noisy GHZ state \eqref{GHZdephased} will produce local full-correlators \cite{noteBG}.
This implies that, no full-correlator inequality (for any number of measurements) is able to detect the nonlocality of state \eqref{GHZdephased} for $N=3$ in the region $p\geq1/2$. Notwithstanding, the nonlocality in this region is successfully detected by the CHSH method.

\subsubsection{Transversal dephasing}
We now analyse the case of the GHZ state~\eqref{GHZ} under dephasing along the transversal $X$ direction. The noisy state is now given by
\begin{equation}
\rho^{x}_{N}=
{\displaystyle\sum\limits_{\substack{k_{i}=0,1\\i=1,\ldots,N}}}
\left(  1-\frac{p}{2}\right)  ^{N-k}\frac{p}{2}^{k} \Pi \left( \left\vert GHZ_{N}^{{\bf k}
}\right\rangle \left\langle GHZ_{N}^{{\bf k}}\right\vert \right)
\label{GHZTdephased}
\end{equation}
with ${\bf k}=(k_1,k_2\ \hdots k_N)$,  $k_i=0$ or $1$, $k={\displaystyle\sum_{i=1,\ldots,N}} k_{i}$, and where $\Pi \left( \left\vert GHZ_{N}^{{\bf k}
}\right\rangle \left\langle GHZ_{N}^{{\bf k}}\right\vert \right)$ stands for the sum of all the $\binom{N}{ k}$ different permutations of $\left\vert GHZ_{N}^{{\bf k}}\right\rangle \doteq X_1^{k_1}\otimes\ldots\otimes X_N^{k_N}\left\vert GHZ_{N}\right\rangle $ with $X_i^0=\openone$. The noisy state \eqref{GHZTdephased} does not have a simple form as~(\ref{GHZdephased}), and the optimal measurements for the MK inequality depend now on both $N$ and $p$. Analytical expressions for the MK violation and the corresponding noise threshold  as functions of $N$ and $p$ are not available. However, using the multipartite CHSH method, a straightforward analysis is possible.

Applying the projector $\left( \ket{+}\bra{+} \right) ^{\otimes N-2}$, with support on all but qubits $i$ and $j$, to \eqref{GHZTdephased} results in the two-qubit state $\rho^{x}_{2}=\left(\left( 1-\frac{p}{2}\right)  ^{2}+\left(  \frac{p}{2}\right)
^{2} \right)\left\vert GHZ_{2}\right\rangle \left\langle GHZ_{2}\right\vert +2\left(
1-\frac{p}{2}\right)  \left(  \frac{p}{2}\right)  (X_i\otimes \openone_j)\left\vert
GHZ_{2}\right\rangle \left\langle GHZ_{2}\right\vert (X_i\otimes \openone_j)$. For this state, one finds $\mathcal{M}^{x}_{\text{CHSH}}=\sqrt{1+(1-p)^{4}}$. The noise threshold obtained is again $p_c^x=1$, independently of $N$,
which reflects the entanglement robustness of GHZ states under transversal local dephasing \cite{Chaves2012a,Chaves2013}.

\subsubsection{General Pauli channels}
An analytical expression for the GHZ state under the general Pauli channel \eqref{Paulimap} can be obtained. Even though the  evolved state is GHZ-diagonal, analytical expressions for the MK violation are again not available. However, the CHSH method offers again a readily calculable bound. One obtains then $\mathcal{M}_{\text{CHSH}}=\sqrt{(p_0+p_1-p_2-p_3)^{2n}+(p_0-p_1-p_2+p_3)^4}$. As a particular interesting case, we  analyse approximate transversal local dephasing defined by $\alpha_1=1-\epsilon$, $\alpha_2=\epsilon/2$ and $\alpha_3=\epsilon/2$. The parameter $\epsilon$ thus measures the deviation of perfect transversal dephasing. In this case, $\mathcal{M}_{\text{CHSH}}=\sqrt{(1-p\epsilon)^2N+(1-p(1-\epsilon/2))^4}$, which, for small values of $p$, can be approximated as $\mathcal{M}_{\text{CHSH}} \approx \sqrt{1+(1-p)^{2N\epsilon}}$, yielding an exponential decay with $N$, as with parallel dephasing, but with the decay rate reduced by a factor $\epsilon$, in a similar fashion to what happens with the entanglement in these noisy states \cite{Chaves2012a,Chaves2013}.

\subsubsection{Non-local content of noisy GHZ states}
To obtain a good lower bound for the local content we use the inequality \eqref{sumCHSH}. For GHZ states \eqref{GHZdephased} under parallel local dephasing, $\rho_1$ and $\rho_2$ are given by
\begin{equation}
\rho_{1,2} = (1-p)^N\ket{GHZ^{\pm}_{2}}\bra{GHZ^{\pm}_{2}}+(1-(1-p)^N) \tilde{\rho}_2^{z}.
\end{equation}
with $\ket{GHZ^{\pm}_{2}}=(1\sqrt{2})\ket{00}\pm\ket{11}$. In this case $p_1=p_2=1/2$. Choosing $A_0=Z$, $A_1=X$, $B_0=\cos{(\theta)}Z+\sin{(\theta)}X$ and $B_1=\cos{(\theta)}Z-\sin{(\theta)}X$ we find the left hand side of \eqref{CHSH1} and \eqref{CHSH2} to be equal to $\cos{(\theta)}+\sin{(\theta)}(1-p)^N$. It is a simple calculation to show that choosing $\theta =\sec^{-1}(\sqrt{1+(1-p)^{2N}})$ the latter value  equals $\mathcal{M}^z_{\text{CHSH}}$.

\begin{figure}[t!]
\centering
\includegraphics[width=\linewidth]{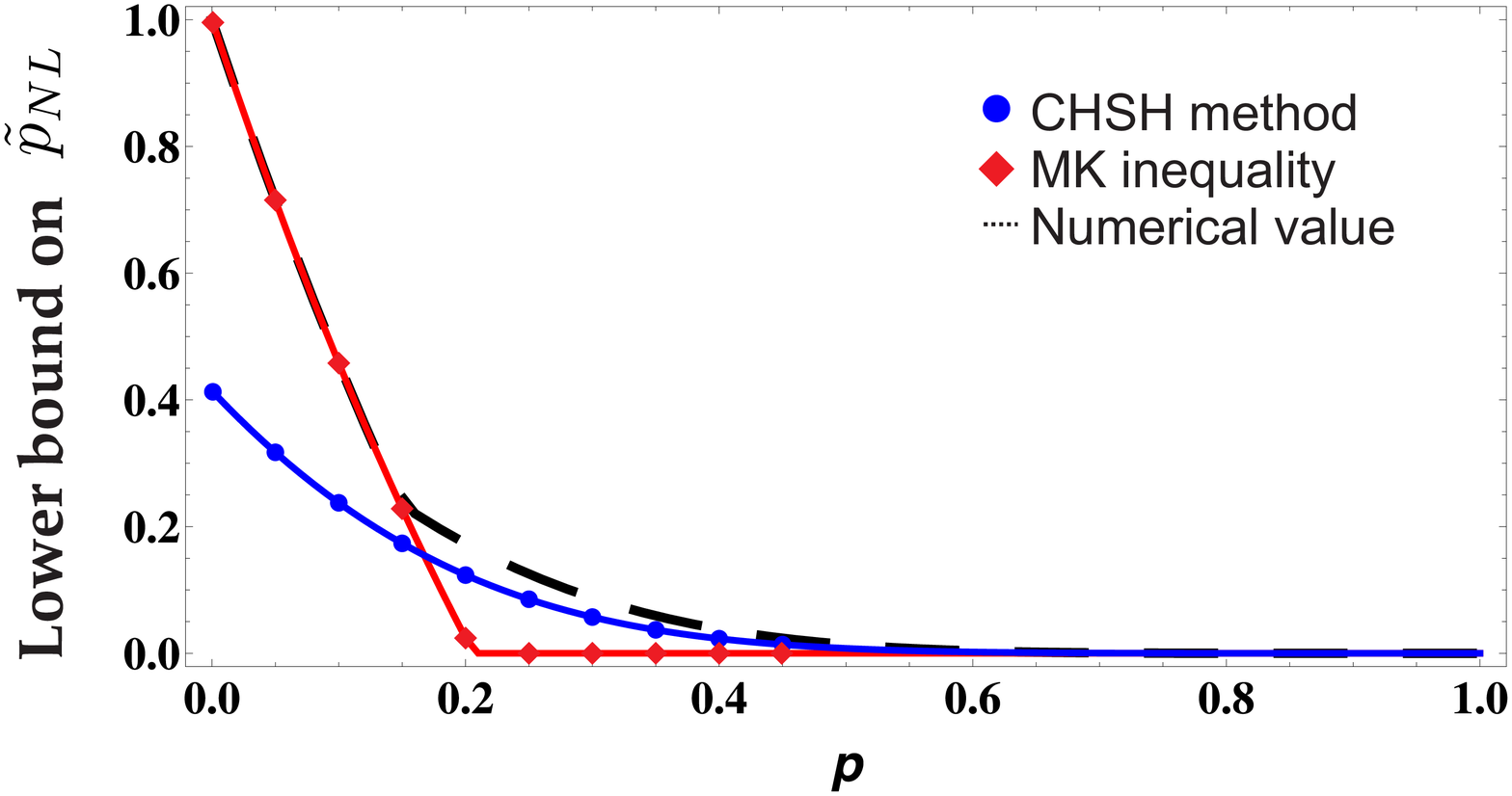}
\caption {
(Color online) Lower bounds on the local content of GHZ state under parallel dephasing, for $N=3$. In red: the bound obtained from the MK inequality \cite{Chaves2012b}; in blue: the new bound \eqref{projpnlGHZ} from the CHSH method; and in black dashed: the value obtained through a numerical optimization described in the main text. For $p > 0.18$, the nonlocal content is better described by the bound from the CHSH method.} \label{fig:NLGHZ}
\end{figure}

So for the GHZ state under parallel dephasing the CHSH method leads to the following lower bound on the nonlocal content
\begin{equation}
    \tilde{p}_{NL}\geq \sqrt{1+(1-p)^{2N}}-1.
    \label{projpnlGHZ}
\end{equation}
In Fig. \ref{fig:NLGHZ}, this bound is compared with the lower bound obtained in Ref. \cite{Chaves2012b} through the MK inequality and with a numerical estimate, for $N=3$. To obtain the numerical estimate we first note that, for $N=3$ and two dichotomic measurements per party, all the facets of local polytope are known, the so called Sliwa inequalities \cite{Sliwa2003}. We have optimized the violation of Sliwa inequalities over all possible projective measurements and using \eqref{lowerbound} obtained the optimal lower bound on ˜$\tilde{p}_{NL}$. As can be seen in Fig. \ref{fig:NLGHZ}, for most of the dynamics, bound \eqref{projpnlGHZ} is tighter than the bound given by the MK inequality.

\begin{figure}[t!]
\centering
\includegraphics[width=\linewidth]{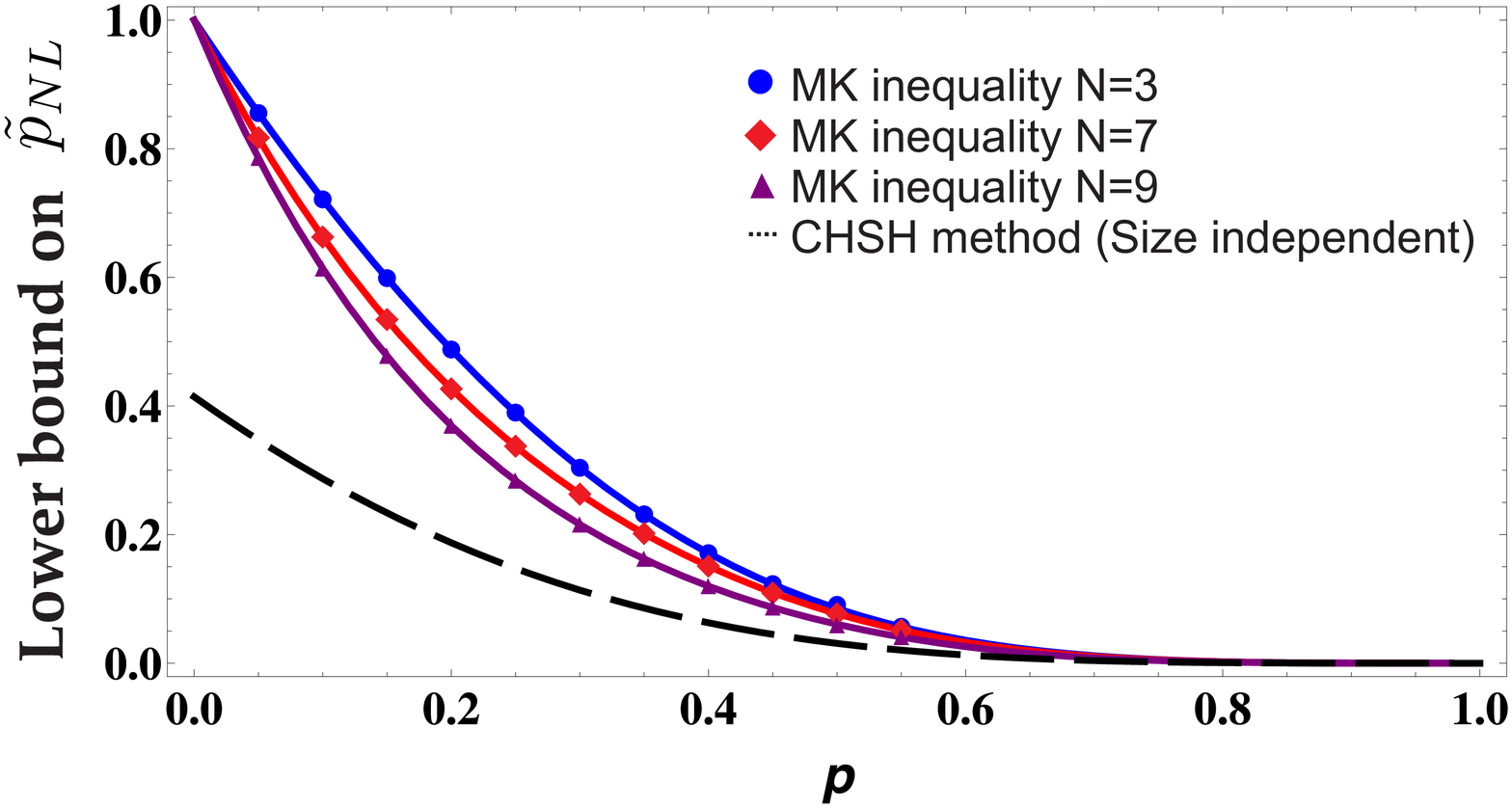}
\caption {(Color online) Lower bounds on the local content of  GHZ state under transversal dephasing. In blue: lower bound from the MK inequality of $N=3$; in red: {\it idem} for $N=7$; in purple: {\it idem} for $N=9$; in black dashed:  lower bound \eqref{projpnlGHZT}  from the CHSH method. The bounds from the MK inequality were obtained through numerical optimisation over all possible projective measurements. The CHSH bound is analytical and independent of $N$.} \label{fig:NLtranserve}
\end{figure}

A similar calculation shows that for GHZ states \eqref{GHZTdephased} under transversal local dephasing, the CHSH method gives
\begin{equation}
    \tilde{p}_{NL}\geq \sqrt{1+(1-p)^{4}}-1.
    \label{projpnlGHZT}
\end{equation}
An analytical expression for the optimal MK violation is not available, as mentioned before. We numerically optimise the MK violation and so derive a numerical lower bound in the nonlocal content, plotted in Fig. \ref{fig:NLtranserve} together with bound \eqref{projpnlGHZT}. The numerical MK bound is tighter, but the required optimization soon becomes unfeasible as $N$ grows. Bound \eqref{projpnlGHZT}, in contrast, is analytical and does not depend on $N$.

\subsection{Noisy W states}
\label{subsec:wstate}
Let us now consider the nonlocality of the noisy W state \eqref{Wnoisy}. We will consider dephasing along the $z$ direction in each of its qubits, which produces the state \cite{Chaves2010}
\begin{equation}
\rho_{W_N}= \frac{1}{N} \left(1-p^{\prime}\right)  \Pi \left(  \ket{00 \ldots 01}\bra{00 \ldots 01} \right)+p^{\prime}  \left\vert W\right\rangle \left\langle W\right\vert,
\label{Wnoisy}
\end{equation}
with $p^{\prime}=\left(  1-p\right)^2$ and $\Pi(.)$ stands for all the permutations. The measurement outcome corresponding to the projector $\ket{0}\bra{0}^{\otimes N-2}$ (that occurs with probability $p=2/N$), produces a two-qubit entangled state of the form $\rho^{2}_{W}$ for which the CHSH violation is $\mathcal{M}_{\text{CHSH}}=\sqrt{1+(1-p)^{4}}$. So we recover the result in Ref. \cite{Laskowski2010} that the dephased W-state is non-local through all the noisy dynamics, that is, $p_c=1$.

Once again we can use the multipartite CHSH method to provide a lower bound to the nonlocal content of this state. However, in this case the project states associated with other measurement outcomes other than $\ket{0}\bra{0}^{\otimes N-2}$ are not local unitarily equivalent to $\rho^{2}_{W}$. Actually they turn out to be separable and given by $\ket{0}\bra{0}^{\otimes N-2}$. Because of that, we must use expression \eqref{neweq} to calculate the lower bound to the local content, which renders
\be
\tilde{p}_{NL}\geq (2\sqrt{1+(1-p)^{4}}-2)/N.
\ee
This bound provides a better estimate for the non-local content of \eqref{Wnoisy} when compared to the one that can be obtained from the Bell inequality used in Ref. \cite{Laskowski2010}. There the inequality used has 
a non-signalling bound that increases exponentially with the number $N$ of qubits, while the violation given by the W-state is approximately independent of $N$. This makes the lower bound decay exponentially while our bound only decays linearly with $N$.


\subsection{Noisy graph-states}

As the last application of the multipartite CHSH method, we study the nonlocality properties of graph states \eqref{graphdef} subject to Pauli channels. In Ref. \cite{Geza2006}, multipartite Bell inequalities specially tailored to detect the nonlocality of graph states have been introduced. For some of these states, these inequalities are violated exponentially in $N$. 
Moreover the violation, for any graph state under any Pauli channel, can be analytically expressed in a compact closed form (see Appendix).

For instance, for graph states under parallel local dephasing, their violation always decreases exponentially fast in $N$, which implies that the associated lower bound on the local content also decreases exponentially with $N$. Nevertheless, it is known that the entanglement in graph states is robust against local noise \cite{Cavalcanti2009,Noisygraph_pra}. With the CHSH method, one easily shows that such entanglement robustness is also reflected in the non-locality robustness.

\begin{figure}[t!]
\centering
\includegraphics[width=\linewidth]{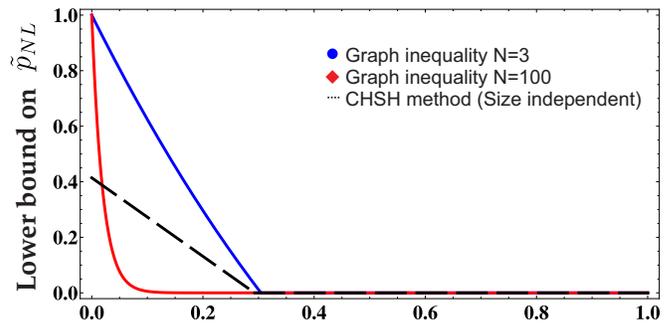}
\caption {
(Color online) Lower bounds on the nonlocal content of for star graph-state of $N$ qubits and $\left\vert I\right\vert=N-1$ (see Appendix for details) and under parallel local dephasing. In red and blue: the bounds obtained from the inequalities of Ref. \cite{Geza2006}, for $N=3$ and $N=100$, respectively; in black dashed: the new bound given by the CHSH method. The CHSH bound is size independent and offers an exponentially tighter estimate as $N$ increases.} \label{fig:graph_nl}
\end{figure}

As an illustration consider a star graph consisting of $N-1$ disconnected qubits connected to one central qubit. One simply projects all but the central qubit and one of the mutually disconnected qubits in the computational basis. Because the projection commutes with the dephasing channel, the two possible final states $\rho_1$ and $\rho_2$ (with $p_1=p_2=1/2$) are also a two-qubit graph state under parallel local dephasing (up to local unitaries). The left hand side of \eqref{CHSH1} and \eqref{CHSH2} are equal to $\mathcal{M}_{\text{CHSH}}=(1-p)\sqrt{2}$. This implies $p_c=1-\sqrt{1/2}$ and the robust (size independent) bound  $\tilde{p}_{NL} \geq \max{ [0,(1-p)\sqrt{2}-1 ]}$. For large $N$ and $p<p_c$, this bound is exponentially tighter than that obtained from the Bell inequalities of \cite{Geza2006}, as shown in Fig. \ref{fig:graph_nl}.

\section{Summary}
\label{sec:conc}

In this work we have used the CHSH inequality in the multipartite scenario, and showed its usefulness to detect the nonlocality of noisy multipartite states. The method consists of locally projecting the multipartite state into a nonlocal two-qubit state that violates the CHSH inequality. We have shown examples of states for which the nonlocality cannot be detected by the W$^3$ZB inequalities consisting only of full-correlators (actually, any full-correlator inequality if $N=3$), but can be detected by the present method. The multipartite CHSH method works well also in situations were it is difficult to analytically find optimal Bell inequality violations, as for GHZ states undergoing transversal dephasing. Furthermore, the method can be easily applied to obtain tight lower bounds to the nonlocal content of correlations.

We believe these findings should contribute to the detection of non-locality for noisy multipartite states. In particular, the present method seems to be the simplest one to experimentally detect  nonlocality in multipartite states.

\begin{acknowledgements}
We thank N. Brunner for motivating us to write this paper and the Benasque Center for Science for hospitality during the Quantum Information Workshop 2013. We also thank the referee for his/her useful comments. This work was supported by the Excellence Initiative of the German Federal and State Governments (Grant ZUK 43) the EU project SIQS, and the National Institute of Science and Technology for Quantum Information, the National Research Foundation and the Ministry of Education of Singapore. LA acknowledges the support from the EU under Marie Curie IEF No 299141.

\end{acknowledgements}

\appendix
\label{sec:append}
\section{Graph states Bell inequalities under Pauli channels}
Given a vertex $i$ of a graph state $\left\vert G_{\bf{0}}\right\rangle $ of $N$ qubits and a subset of its neighbors $I\subseteq\mathcal{N}(i)$, such that none of the vertices in $I$ are connected by and edge, the Bell operator $\mathcal{B}(i,I)=K_{i} {\displaystyle\prod\limits_{j\epsilon I}} (1+K_{j})$ ($K_{i}=X_{i} \prod_{j \in \mathcal{N}_{i}} Z_{j}$ are the generators of the graph, that is, $K_i  \ket{G_{\bf{0}}} =  \ket{G_{\bf{0}}}$) defines a Bell inequality given by \cite{Geza2006}
\begin{equation}
\left\vert \left\langle \mathcal{B}\right\rangle \right\vert =\left\vert
\left\langle \mathcal{B}(i,I)\right\rangle \right\vert \leq L\left(
\left\vert I\right\vert +1\right)  ,\label{graph inequality}%
\end{equation}
with a classical bound given by $L(m)=2^{\left(  m-1\right)  /2}$ for $m$ odd, and $L(m)=2^{m/2}$ for $m$ even. The inequality is maximally violated by the graph $\left\vert G_{\bf{0}}\right\rangle
$ with $\left\langle \mathcal{B}(i,I)\right\rangle =2^{\left\vert I\right\vert }$.
\par Under the action of a Pauli map, the graph state will turn into a graph diagonal mixed state $\rho_{G}=\sum p_{\mu}\ket{G_{\mu}}\bra{G_{\mu}}$, with $\ket{G_{\mu}}=Z^{\mu_1} \otimes Z^{\mu_2} \dots \otimes Z^{\mu_N}\ket{G_{\bf{0}}}$, where $\mu=(\mu_1,\dots,\mu_N)$ is a multi-index, $\mu_j$ can assume values $0$ or $1$ and the weights $p_{\mu}$ depend on the exact form of the Pauli map.
The expectation value of the Bell operator $\mathcal{B}(i,I)$ on this state is given by
\begin{align}
\label{vio_graph_dia}
\left\langle \mathcal{B}\right\rangle_{\rho_{G}}  &  = \left\langle \sum p_{\mu}
K_{i}\displaystyle\prod\limits_{j\in I}(1+K_{j}) \ket{G_{\mu}}\bra{G_{\mu}} \right\rangle \\ \nonumber
& = \sum p_{\mu} (-1)^{\mu_i}\displaystyle\prod\limits_{j\in I}(1+(-1)^{\mu_j}),
\end{align}
where we have used that $K_i  \ket{G_{\mu}} =  (-1)^{\mu_i}\ket{G_{\mu}}$. From \eqref{vio_graph_dia} it follows that the only terms in the convex sum $\rho_{G}=\sum p_{\mu}\ket{G_{\mu}}\bra{G_{\mu}}$ contributing to the expectation of the Bell operator are $\mu^{0}=(0,0,\dots,0)$ and $\mu^{1}=(1,0,\dots,0)$, that is, $\left\langle \mathcal{B} \right\rangle_{\rho_{G}}= (p_{\mu^{0}}-p_{\mu^{1}})2^{\vert I\vert}$.
\par As a matter of fact, consider any graph state undergoing local dephasing. From \eqref{vio_graph_dia} we see that $\left\langle \mathcal{B}\right\rangle_{\rho_{G}}= (1-p)(  1-p/2)^{N-1}2^{\vert I\vert}$, which shows an exponential decay in the violation with $N$, that is also reflected in a exponential decay of the associated lower bound for the nonlocal content (see Fig. \ref{fig:graph_nl}).

\bibliography{projections}

\begin{thebibliography}{48}
\bibitem [{not({\natexlab{a}})}]{noteoptimal}%
  \BibitemOpen
  \href@noop {} {} ({\natexlab{a}}),\ \bibinfo {note} {in general we are
  interested in the maximal violation of CHSH inequality that can be obtained
  by the projection of $N-2$ qubits. Note that the projected two-qubit state
  can, in principle, be explicitly written as a function of the $2(N-2)$ angles
  and the $N-2$ dichotomic parameters describing, respectively, the projectors
  and its outcomes. One can then maximize the CHSH violations (via the
  criterion in Ref. \cite{HHH95}) optimizing over these variables. For example,
  for GHZ states under parallel and transversal dephasing we have numerically
  observed (up to $N=5$) that the projections in the $X$ basis are optimal. We
  have considered this projective measurement for arbitrary $N$, which provides
  a lower bound on the critical noise strengh $p_c$ and the nonlocal weight
  $\tilde{p}_{NL}$.}\BibitemShut {Stop}%
\end{thebibliography}

\end{document}